\documentclass[aps,nofootinbib,12pt]{revtex4}
\usepackage{amsmath}
\usepackage{amsfonts}
\usepackage{amssymb}
\usepackage{color}
\usepackage{graphicx}
\usepackage{mathrsfs}

\def\bc{\begin{center}}
\def\nno{\nonumber}
\def\ec{\end{center}}
\def\be{\begin{eqnarray}}
\def\ee{\end{eqnarray}}

\definecolor{dyellow}{rgb}{1.,0.8,.0}
\definecolor{myblue}{rgb}{.1,.1,.7}
\definecolor{dcyan}{rgb}{.0,.6,.6}
\definecolor{dmagenta}{rgb}{0.6,0.0,0.6}
\definecolor{brown}{rgb}{0.6,0.2,0.}
\definecolor{darkblue}{rgb}{.0,.0,0.5}
\definecolor{darkred}{rgb}{0.75,0.0,0.0}
\definecolor{orange}{rgb}{1.,.6,.0}
\definecolor{dorange}{rgb}{0.8,.4,.0}
\definecolor{darkgreen}{rgb}{0.0,0.6,0.0}
\definecolor{purple}{rgb}{.4,.0,.4}
\definecolor{lightgrey}{rgb}{0.7, 0.7, 0.7}
\definecolor{grey}{rgb}{0.4, 0.4, 0.4}

\def\al{\alpha}

\def\ka{\kappa}

\def\si{\sigma}
\def\om{\omega}

\def\pa{\partial}

\def\Ga{\Gamma}

\def\Si{\Sigma}
\def\Om{\Omega}

\def\h{\hat}
\def\bpsi{\bar{\Psi}}

\begin{document}

\title{Quasinormal Modes of Charged Fermions and Phase Transition of Black Holes} \vskip 2cm \vskip 2cm
\author{Rong-Gen Cai$^1$}\email{cairg@itp.ac.cn}
\author{Zhang-Yu Nie$^1$}\email{niezy@itp.ac.cn}
\author{Bin Wang$^2$}\email{wangb@fudan.edu.cn}
\author{Hai-Qing Zhang$^1$}\email{hqzhang@itp.ac.cn}
\address{$^1$Key Laboratory of Frontiers in Theoretical Physics,
Institute of Theoretical Physics, Chinese Academy of Sciences,
   P.O. Box 2735, Beijing 100190, China}
\address{$^2$Department of Physics, Fudan University, Shanghai 200433, China}


\begin{abstract}
We study the quasinormal modes of massless charged fermions in a
Reissner-Nordstr\"om-anti-de Sitter black hole spacetime. In the
probe limit, we find that the imaginary part of quasinormal
frequency will become positive when the temperature of the black
hole is below a critical value. This indicates an instability of the
black hole occurs and a phase transition happens. In the AdS/CFT
correspondence, this transition can be viewed as a superconducting
phase transition and the bulk fermion is regarded as the order
parameter.  When the coupling of the fermions and the background
electric field becomes stronger, the critical temperature of the
phase transition becomes higher.  If the interaction between the
fermion and the electric field can be ignored, namely in the case of
a neutral fermion, the imaginary part of the quasinormal modes is
always negative, which indicates that the black hole is stable and
no phase transition occurs.

\end{abstract}


\maketitle

\section{Introduction}

Over the past years the holographic models of superconductors have attracted a lot of attentions since the works~\cite{gub,hart}. For reviews see
\cite{reviews}. In a simple model with the Einstein-Maxwell-complex scalar field system with a negative cosmological constant, the charged scalar
field will condensate when the temperature of the anti-de Sitter (AdS) black hole is below a critical value. Above the critical temperature, the
system has the Reissner-Norstr\"om (RN)-AdS black hole solution with a trivial scalar field, while below the critical temperature, a hairy black
hole solution is more stable with a nontrivial scalar field. This indicates that a phase transition happens between the RN-AdS black hole and a
hairy black hole when the temperature of the black hole arrives at the critical value. According to the AdS/CFT correspondence~\cite{maldacena,
polyakov, witten}, the phase transition of the AdS black hole can be mapped to a superconducting phase transition on the boundary of the AdS
space~\cite{hart}, the condensation of charged scalar field around the black hole corresponds to a condensation of the charged operators in the
boundary field theory.

It is well known that a neutral scalar field perturbation of an
asymptotically AdS spacetime is stable if the mass of the scalar
field satisfies the Breitenlohner-Freedman (BF) bound~\cite{bf}.
However, the perturbation of charged scalar field in an
asymptotically AdS spacetime can cause the background to become
unstable~\cite{gub, meada,murata,kono}. In the holographic
superconductor models, the presence of the instability of the
perturbation of the charged scalar field just indicates the
occurrence of the condensation in the boundary field theory.

To reveal the stability of a black hole, a useful method is to study the quasinormal modes of some perturbations in the black hole background,
for a review see \cite{kok}. Recently it was found that quasinormal modes of charged scalar field in some AdS black hole backgrounds can be used
to disclose the relation to the superconductor transition ~\cite{he}. The occurrence of the unstable modes was observed in consistent with the
superconducting phase transition.

Besides the charged scalar field, it is of great interest to examine whether some other fields can experience the condensation in the AdS black
hole background, for example the charged fermionic field, its condensation might be related to the model of color superconductor in QCD theory.
The perturbation of a charged fermionic field in an asymptotically flat spacetime was found always with decay modes~\cite{yau}. The fermionic
field will either be absorbed by the horizon or go to the spatial infinity and cannot condensate near the black hole. It is well known that the
AdS spacetime has different asymptotical behavior from that of the asymptotically flat spacetime and this difference plays crucial role on the
dynamics of the inner geometry. For example, the late-time tail of the perturbations with a power law form in an asymptotically flat spacetime
gives way to the exponential form  in an asymptotically AdS spacetime ~\cite{horo,wang}. In this work we will concentrate on the study of the
charged fermonic field perturbation in the background of an AdS black hole.

We employ a simple RN-AdS spacetime geometry while leaving fermionic
fields as probes. We find that when the temperature is below a
certain critical value, the imaginary part of the quasinormal modes
will become positive, which indicates the occurrence of the
instability of the black hole. This shows that at the critical
temperature there appears a phase transition, the RN-AdS black hole
background becomes unstable and a new stable hairy black hole
solution is expected to emerge. We observe that the critical
temperature grows with the increase of the charge coupling constant.
This means that the stronger the fermions couple to the Maxwell
field, the easier the instability occurs. Our result is consistent
with the holographic superconducting phase transition discussed in
\cite{reviews}\cite{he}, however in our result the order parameter
is a charged fermion instead of the charged scalar field. In this
sense our model might be regarded as a simple model of color
superconductor in QCD theory.

This paper is organized as follows. In Sec.~\ref{sect:act} we briefly introduce the RN-AdS geometry and the action of the charged fermions in
this background. Under an explicit representation, the Dirac equation is splitted in Sec.~\ref{sect:split} to a concise form. In
Sec.~\ref{sect:qnm}, we use the Horowitz-Hubeny approach to Fourier expand the Dirac equations for the convenience to  study the quasinormal
modes of the fermionic perturbation numerically. We give the numerical results and analysis of the quasinormal modes in Sec.~\ref{sect:num}. We
conclude our paper in Sec.~\ref{sect:con}.

\section{The Background Geometry and  Action}
\label{sect:act}

  For a general $d$-dimensional RN-AdS spacetime, the
  metric is
 \be\label{metric} &&ds^2=-f(r)dt^2+\frac{dr^2}{f(r)}+r^2d\Om^2_{d-2,k},\\
 \text{where \quad}&&f(r)=k+\frac{r^2}{L^2}+\frac{Q^2}{4r^{2d-6}}-(\frac{r_0}{r})^{d-3}.\ee
 $L$ is the AdS radius, $Q$ is the charge of the black hole and $r_0$ is related to the black hole mass
 $M$  via
  \be r_0^{d-3}=\frac{16\pi G_d M}{(d-2)A_{d-2}},\ee
  where $A_{d-2}=2\pi^{\frac{d-1}{2}}/\Ga(\frac{d-1}{2})$ is the
  area of a unit $(d-2)$-sphere, $G_d$ is the Newton gravitational
  constant in the $d$-dimensional spacetime. $d\Om^2_{d-2,k}$ in
  (\ref{metric}) is the metric of constant curvature. If $k=0$,
  it is the metric of a flat Euclidean space
  $\bf{R}^{d-2}$; if $k>0$, it is the line element of a
  $(d-2)$-sphere with radius $\frac{1}{\sqrt{k}}$; and if $k<0$, it is the metric of
  a hyperbolic plane with radius of curvature $\frac{1}{\sqrt{-k}}$.
  Without loss of generality, one can take $k=0,$ and $\pm 1$.

We will  focus on $k=0$ case in this paper for simplicity and also for the relation to a superconductor in a plane. For the cases with $k\ne 0$,
similar results can be obtained.   In particular, we will consider a $4$-dimensional RN-AdS spacetime and let $G_d=1$
  \be\label{metric2}
  &&ds^2=-f(r)dt^2+\frac{dr^2}{f(r)}+r^2(dx^2+dy^2),\ee
 The temperature of the black hole now is
 \be\label{temp} T_H=\frac{3r_+}{4\pi L^2}-\frac{Q^2}{16\pi r_+^{3}}.\ee
where $r_+$ is the horizon radius of the black hole.

 The spin connection is defined as:
 \be \om_{\h a\h bc}=e_{\h ad}\pa_ce^d_{\h b}+e_{\h ad}e^f_{\h
 b}\Ga^d_{\ fc},\ee
 where $e^{\h a}_b$ is the tetrad~\footnote{The un-hatted letter $b$ is the index of the background spacetime
while the hatted letter $\h a$ denotes the index of the tangent
space.}, $\Ga^a_{\ bc}$ is the
 Christoffel connection and $\om_{\h a\h bc}=-\om_{\h b\h a c}$. The nonvanishing spin connections $\om_{\h a\h bc}$ for the metric (\ref{metric2}) are:
 \be \om_{\h r\h t t}=-\om_{\h t\h rt}=\frac1 2 f',\ \ \om_{\h x\h rx}=-\om_{\h r\h xx}=\sqrt{f},\ \ \om_{\h
 y\h ry}=-\om_{\h r\h yy}=\sqrt{f}.\ee
 where a prime denotes the derivative with respect to $r$.

We consider the Einstein-Maxwell-fermion system with the action
 \be
 \label{action}
 S_{T}=S_{g}+S_{m},\ee
 where
  \be S_{g}&=&\frac{1}{2\ka_4^2}\int d^4x
 \sqrt{-g}\left( \mathcal{R}-\frac{6}{L^2}\right ),\\
 S_{m}&=&\mathcal{N}\int d^4x\sqrt{-g}\left (-\frac{1}
 {4}F_{ab}F^{ab}+i\big(\bar{\Psi}\Ga^a(D_a-iqA_a)\Psi-m\bpsi\Psi\big) \right ),\ee
 $\ka_4$ is the 4-dimensional gravitational constant, $\mathcal{R}$ is  Ricci scalar,
 $\mathcal{N}$ is a total coefficient of the action of matter,
  $q$ is the coupling constant between the fermion field and Maxwell field. $\Ga^a$ is the Dirac gamma matrix
 and
 \be D_c&=&\pa_c+\frac1 2 \om_{\h a\h bc}\Si^{\h a\h b},\qquad \Si^{\h
 a\h b}=\frac1 4[\Ga^{\h a},\Ga^{\h b}],\nno\\
 \Ga^b&=&e^b_{\h a}\Ga^{\h a}.\ee
The Dirac equation for the fermion  is
   \be\label{de} \Ga^a(D_a-iqA_a)\Psi-m\Psi=0.\ee
In the action (\ref{action}), clearly we have a simple RN-AdS black
hole solution with the electric potential $A_t$ and a vanishing
fermion $\Psi$:
    \be A_t=Q(\frac{1}{r}-\frac{1}{r_+}),\quad \Psi=0.\ee
 In the following, we will
work in the probe limit which means that the fermionic field does not backreact on the metric and Maxwell field.

\section{The Split of the Dirac Equation}
\label{sect:split}

We can write the wave function $\Psi(r,x_{\mu})$ into the momentum
space
  \be\label{so}
 \Psi(r,x_{\mu})=\psi(r)e^{-i\om
 t+i\vec{k}\cdot \vec{x}}.\ee
where $x_{\mu}$ denotes the coordinates in the boundary,
$x_{\mu}=(t,x,y)$, while $\vec{k}=(k_x,k_y)$ and $\vec{x}=(x,y)$.
Under this transformation, we can write the Dirac equation
(\ref{de}) into
 \be \label{de2}\sqrt{f}\Ga^{\h r}\pa_r\psi-\frac{i\om}{\sqrt{f}}\Ga^{\h
 t}\psi+\frac{i\vec{k}\cdot \Ga^{\hat{\vec{x}}}}{r}\psi+\frac1
 4[\frac{f'}{\sqrt{f}}+\frac{4\sqrt{f}}{r}]\Ga^{\h
 r}\psi-(iq\Ga^aA_a+m)\psi=0.\ee
where $\vec{k}\cdot \Ga^{\hat{\vec{x}}}=k_x\Ga^{\h x}+k_y\Ga^{\h
y}$.  We can choose the Dirac gamma matrices as \cite{lee}
 \be \Ga^{\h t}=\left(
                  \begin{array}{cc}
                    0 & - I\\
                    I & 0 \\
                  \end{array}
                \right),
     \Ga^{\hat{i}}=\left(
                       \begin{array}{cc}
                         0 & \si^{\h i} \\
                         \si^{\h i} & 0 \\
                       \end{array}
                     \right)\ee
where $I$ is the $2\times2$ unit matrix,  $\si^{\h i}$ is the Pauli
matrix, explicitly,
 \be \si^{\h x}=\left(
       \begin{array}{cc}
         0 & 1 \\
        1 & 0 \\
       \end{array}
     \right),\quad \si^{\h y}=\left(
                           \begin{array}{cc}
                             0 & -i \\
                             i & 0 \\
                           \end{array}
                         \right),\quad \si^{\h r}=\left(
                                               \begin{array}{cc}
                                                 1 & 0 \\
                                                 0 & -1 \\
                                               \end{array}
                                             \right). \ee
It is easy to decompose the fermion filed $\Psi$ into $\Psi_+$ and
$\Psi_-$, which are the eigenvectors of $\Ga^5$, {\it i.e.}
                                             \be\Psi=\left(
                                                \begin{array}{c}
                                                  \Psi_+ \\
                                                  \Psi_- \\
                                                \end{array}
                                              \right),\quad P_{\pm}\Psi=\pm\Psi_{\pm},\quad P_{\pm}
                                              =1\pm\Gamma^5,\quad \Ga^5=i
\Ga^t\Ga^x\Ga^y\Ga^r.\ee %
Under this representation, the Dirac equation (\ref{de2}) can be
decomposed into %
\be \label{decom1}&&(\sqrt{f}\pa_r+\frac1
4\frac{f'}{\sqrt{f}}+\frac{\sqrt{f}}{r})\si^{\h r}\psi_-+\frac i r
(\vec{k}\cdot \vec{\si})\psi_-+\frac{i}{\sqrt{f}}(\om+qA_t)\psi_--m\psi_+=0,\\
\label{decom2}&&(\sqrt{f}\pa_r+\frac1
4\frac{f'}{\sqrt{f}}+\frac{\sqrt{f}}{r})\si^{\h r}\psi_++\frac i r
(\vec{k}\cdot
\vec{\si})\psi_+-\frac{i}{\sqrt{f}}(\om+qA_t)\psi_+-m\psi_-=0.\ee It
is easy to see from eqs.(\ref{decom1}) and (\ref{decom2}) that if
$\om\rightarrow-\om$ and $q\rightarrow-q$, there is a permutation
symmetry  $\psi_+\leftrightarrow \psi_-$ in (\ref{decom1}) and
(\ref{decom2}). In the following, we will take $m=0$, because the
chiral modes of $\psi$ will decouple in this case, which can be
easily seen from eqs.~(\ref{decom1}) and (\ref{decom2}).

For the massless fermions, we will  focus on the modes of $\psi_+$ , because the modes of $\psi_-$  can be obtained accordingly if we change the
sign of $\om$ and $q$. Furthermore, for simplicity, we can set $k_y=0$ because of the symmetry of $(\vec{x},\vec y)$-plane \cite{liu}. We rewrite
$\psi_+$ as
 \be \psi_+=r^{-1}f^{-1/4}\tilde{\psi}.\ee
And then eq.~(\ref{decom2}) can be further simplified into
 \be\label{de3} \si^{\h r}\tilde{\psi}'-\frac i
 f(\om+qA_t-\frac{\sqrt{f}}{r}k_x\si^{\h x})\tilde{\psi}=0.\ee
Here $\tilde{\psi}$ is a 2-component fermion wavefunction, we can
decompose it as $\tilde{\psi}=\left(
                \begin{array}{c}
                  \psi_1 \\
                  \psi_2 \\
                \end{array}
              \right)
$ under the $\si^{\h r}$ matrix. Under this representation, eq.~(\ref{de3}) can be further decomposed into
 \be\label{decom3} &&\psi_1'-\frac i f
 (\om+qA_t)\psi_1+\frac{i}{r\sqrt{f}}k_x\psi_2=0,\\
\label{decom4}&&\psi_2'+\frac i f
 (\om+qA_t)\psi_2-\frac{i}{r\sqrt{f}}k_x\psi_1=0.\ee
Like in the case of eqs.~(\ref{decom1}) and (\ref{decom2}), if
$\om\rightarrow-\om, q\rightarrow-q$ and $k_x\rightarrow-k_x$, there
is also a permutation symmetry $\psi_1\leftrightarrow \psi_2$
between eqs. (\ref{decom3}) and (\ref{decom4}). Therefore, we can
only focus on the modes $\psi_1$ by extracting $\psi_2$ from
eq.(\ref{decom3}) and then inserting it into eq.(\ref{decom4}).

\section{The Quasinormal Modes}
\label{sect:qnm}

To calculate the quasinormal modes of the charged fermion, we
introduce the tortoise coordinates $r_*$ by
 \be \label{tor} dr_*=\frac{dr}{f}.\ee
 Now, the black hole horizon is located at $r_*\rightarrow-\infty$ and
 the spatial infinity is at a finite value of $r_*$.

The ingoing boundary conditions at the horizon requires the wavefunction to have the form
 \be \label{ingoing} \psi_1(r)=e^{-i\om r_*}u(r).\ee
For simplicity of the numerical calculation, we define $x =1/r$. Thus, the black hole horizon is at $x_+=1/r_+$ while the infinite boundary
$r\rightarrow\infty$ locates at $x=0$. And the equation of $u$ is given as
 \be\label{de4}
 S(x)\ddot{u}(x)+\frac{T(x)}{x-x_+}\dot{u}(x)+\frac{U(x)}{(x-x_+)^2}u(x)=0,\ee
 where the overdot denotes the derivative with respect to $x$ and
 \be \label{S}S(x)&=&\left(-L^2 Q^2 x^3 x_+^3+4 x_+^2+4 x x_++4 x^2\right){}^2,\\
     \label{T}T(x)&=&\frac{1}{2} \left(L^2 Q^2 x^3 x_+^3-4 x_+^2-4 x x_+-4
   x^2\right) \left(4 L^2 Q^2 x^3 x_+^3+16 i L^2 \omega
   x_+^3-3 x^2 \left(L^2 Q^2 x_+^4+4\right)\right),\nno\\ \\
    \label{U}U(x)&=&2 L^2 x_+^3 \left\{2 \omega  \left[-8 L^2 q Q x_+^4-4 i L^2
   Q^2 x^3 x_+^3+8 L^2 q Q x x_+^3+3 i x^2 \left(L^2 Q^2
   x_+^4+4\right)\right]\right.\nno\\&&\left.-\left(x-x_+\right) \left[2
   \left(L^2 Q^2 x^3 x_+^3-4 x_+^2-4 x x_+-4 x^2\right)
   k_x^2+i q Q \left(-8 i L^2 q Q x_+^4\right.\right.\right.\nno\\&&\left.\left.\left.+2 L^2 Q^2 x^3
   x_+^3+8 x_+^2+8 x \left(i L^2 q Q x_+^3+x_+\right)-x^2
   \left(3 L^2 Q^2 x_+^4+4\right)\right)\right]\right\}. \ee
Although the complex number $i$ occurs in the potential $U(x)$ here,
 it was argued in~\cite{lea, kok2, cho, sim, jing} and references
therein that one still can have correct quasinormal modes in the
cases with such a complex potential if some proper boundary
conditions are imposed.

Next we employ the Horowitz-Hubeny method~\cite{horo}
 to calculate the quasinormal modes of this charged fermion.  $S(x),
  T(x)$ and $U(x)$ can be Fourier expanded near $x_+$ to a finite
  term, {\it i.e.},
  $S(x)=\sum_{n=0}^{6}s_n(x-x_+)^n,
  T(x)=\sum_{n=0}^{6}t_n(x-x_+)^n$ and
  $U(x)=\sum_{n=0}^{4}u_n(x-x_+)^n$. We also expand the solution
  $u(x)$ around $x_+$ as
  \be\label{expan} u(x)=(x-x_+)^{\al}\sum_{n=0}^{\infty}a_n(x-x_+)^n.\ee
At the leading order,  the equation has two solutions $\al=1/2$ and
$\al=\frac{8 i L^2  x_+}{12-L^2 Q^2 x_+^4}\om$. It is obvious that
the solution $\al=\frac{8 i L^2  x_+}{12-L^2 Q^2 x_+^4}\om$ gives an
outgoing mode at the horizon. As a result, we choose $\al=1/2$ in
the expansion (\ref{expan}). Substituting eq.(\ref{expan}) into the
eq.(\ref{de4}), we can obtain a recursion relation \cite{jing}
 \be
 a_n=-\frac{1}{P_n}\sum_{k=0}^{n-1}[(k+\al)(k+\al-1)s_{n-k}+(k+\al)t_{n-k}+u_{n-k}]a_k,\ee
 where $P_n=(n+\al)(n+\al-1)s_0+(n+\al)t_0+u_0$.

In the numerical calculations, we set the initial data $a_0=1$ whose
scaling will not affect the final quasinormal  modes $\om$ because
the equation of $\psi$ is linear. In practice, we can solve the
expansion up to a large number $n=N$, then compare the results got
from $N$ and the results from $n>N$, if the error of the two results
is of the order $10^{-5}$, we will believe that the modes $\om_N$ is
acceptable.

\section{Numerical Results}
\label{sect:num}

 The boundary condition on the spatial infinity is that the fermion field vanishes there, which means
  \be u(0)=\sum_{n=0}^{N}a_n(0-x_+)^{n+\al}=0.\ee
 This equation gives the eigenvalues of $\om$.  The quasinormal modes can be decomposed into
  real and imaginary parts:
 \be \om= \text{Re}(\om)+ i \text{Im}(\om).\ee
If Im$(\om)<0$, the perturbations always decay exponentially and
then the background black hole is stable; if Im$(\om)>0$, however,
the modes grow exponentially, which means the perturbations will
make the background black hole unstable. This instability suggests a
phase transition of black holes~\cite{meada, he}.

Following \cite{gub}, in the calculation we set $r_+=Q=1$. In this case, changing $L$ corresponds to the change of the temperature of the black
hole. Furthermore, we set the x-direction momentum $k_x=1$. For the calculation precision we Fourier expand the equation up to $N=300$.

\begin{figure}[htb]
\includegraphics[scale=0.6]{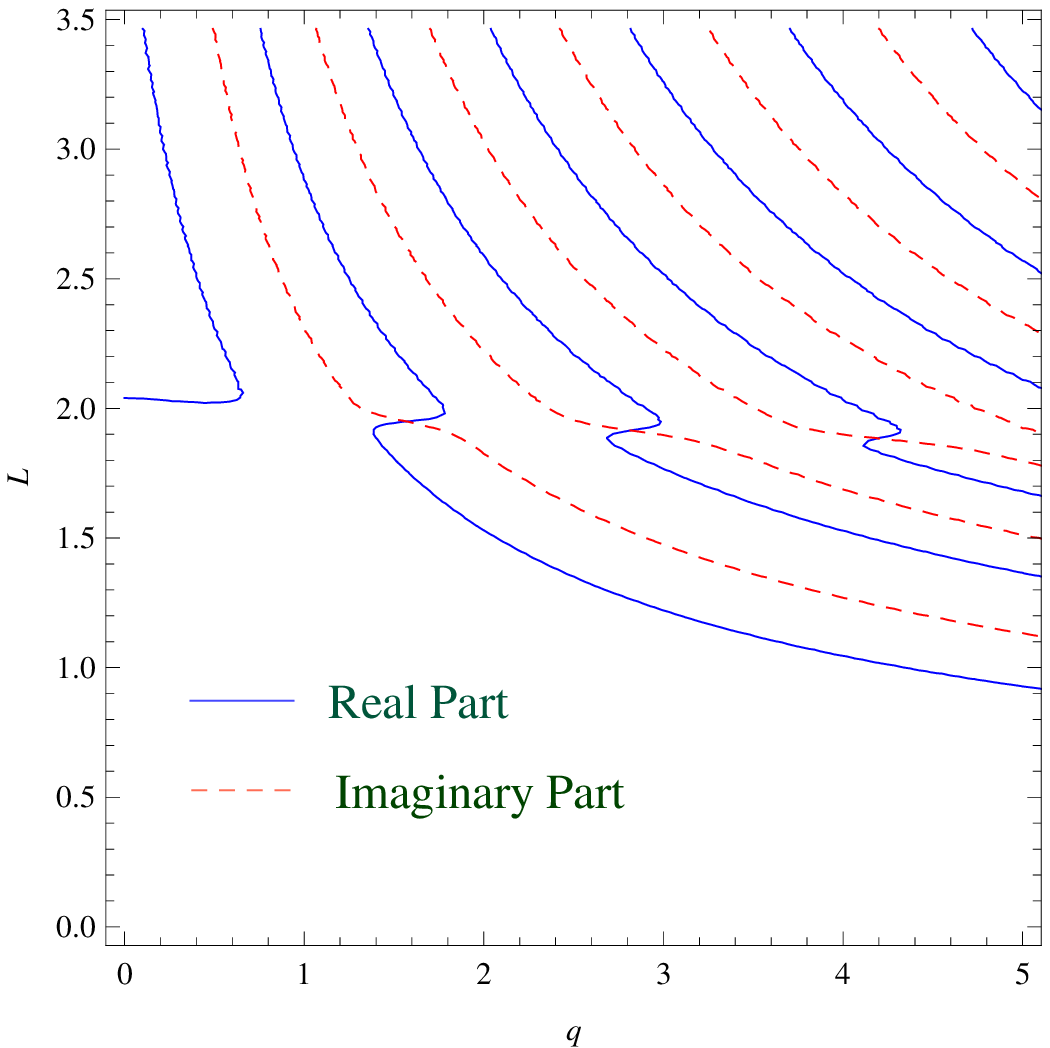}
\includegraphics[scale=0.76]{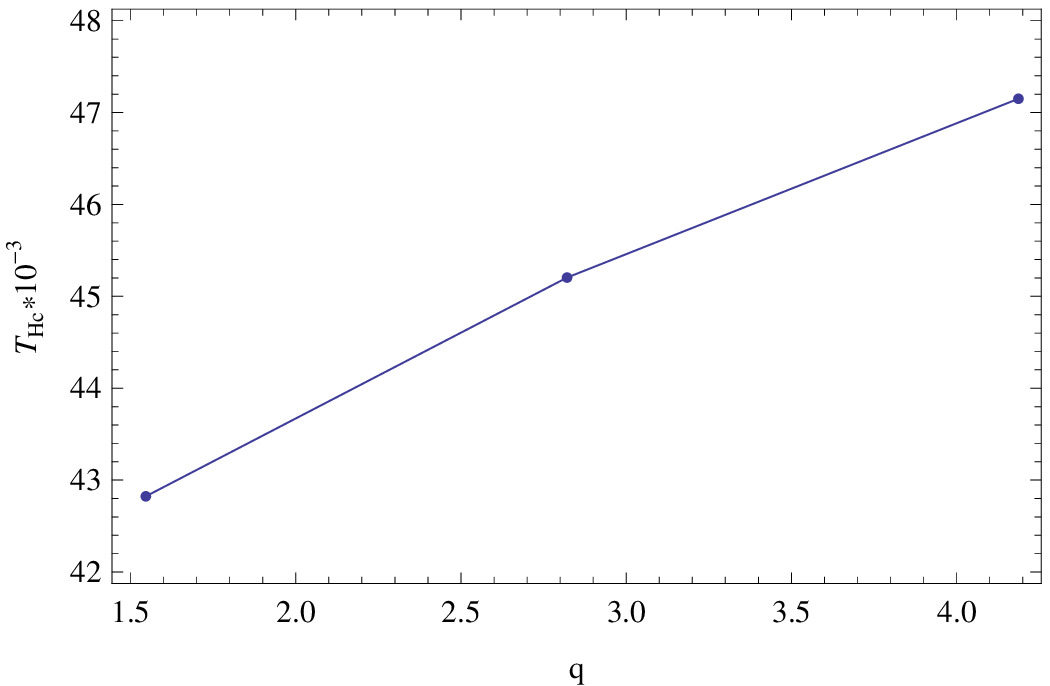}
\caption{\label{qL} (Left Panel) Eq.~(\ref{de4}) with $\om=0$ gives
the two constraints on $q$ and $L$: the real part (blue) and
imaginary part (red  dashed). (Right Panel) The critical temperature
$T_{Hc} $versus $q$.}
\end{figure}

To find the marginally stable perturbation where $\Psi$ depends only
on $r$ and is infinitesimally small, taking $\om=0$ in (\ref{de4}),
we will have an equation on $q$ and $L$ from eq.~(\ref{de4}).
Requiring both the real and imaginary parts of the resulting
equation on $q$ and $L$ to vanish, we have two constraints on the
relation between $q$ and $L$, whose numerical results are shown in
the left panel of Fig.\ref{qL}. It indicates that only $q$ and $L$
at the crossing points of the two constraint curves can give the
marginally stable solution with $\om=0$. The three crossing points
$(q,L)$ shown in the left panel of Fig.\ref{qL} are $(1.546, 1.951),
(2.821, 1.915)$, and $(4.187, 1.887)$, respectively.  When $q$
increases, we see that $L$ at the crossing points decreases.
According to the temperature formula (\ref{temp}) of the black hole,
the temperature increases as $L$ decreases.  Thus the stronger the
fermion field couples to the Maxwell field, the easier the
instability occurs. This is physically reasonable. We show this
relation of the critical temperature and the coupling $q$ in the
right panel of Fig.\ref{qL}.

\begin{figure}[htb]
\includegraphics[scale=0.76]{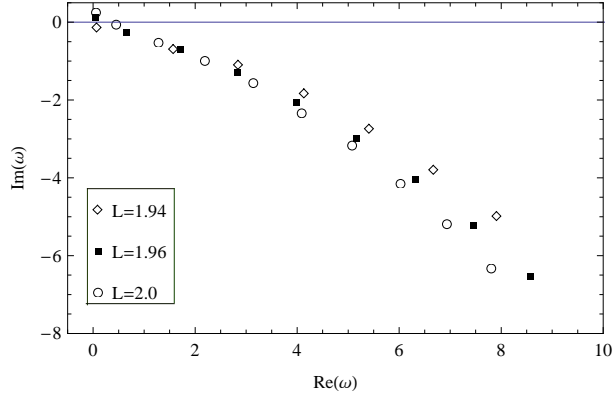}
\caption{\label{reim} The real and imaginary parts of the modes when
changing $L$ while $q$ is fixed as $q=1.546$.}
\end{figure}

The marginally stable perturbation cannot guarantee the existence of the instability but just highly suggestive \cite{gub}. In examining the
stability of the perturbation, one can release the condition $\om=0$, then in principle there does not exist any constraint relation between $q$
and $L$. In this paper we are interested in the quasinormal modes of the fermion perturbation and expect to see whether a positive imaginary part
of quasinormal frequency which indicates the instability of the fermion perturbation will appear. Fig.\ref{reim} shows the frequencies of the
quasinormal modes in the case with a fixed $q=1.546$ and different $L$. It can be seen that when $L>L_c=1.951$ the imaginary part of the
quasinormal frequency becomes positive, which means that the black hole is unstable under the perturbation when the temperature is low enough.
$L=L_c$ corresponds to the critical temperature of the black hole $T_{Hc}$, when $T>T_{Hc}$ the RN-AdS black hole is stable under the fermonic
perturbation, while the RN-AdS black hole becomes unstable when $T<T_{Hc}$ and the fermonic hair is expected to condensate and attach to the
black hole. The RN-AdS black hole will give way to a fermion haired black hole.

\begin{table}
\caption{\label{tablereim} Real and imaginary parts of the
quasinormal modes shown in Fig.\ref{reim}.}
\begin{center}
\begin{tabular}{|c|c|c|c|c|c|c|c|c|c|c|c|}
  \hline
   \multicolumn{2}{|c|}{Node}& 1st & 2nd & 3rd & 4th & 5th & 6th & 7th & 8th & 9th&10th \\
  \hline
   $L=2$ & Re($\om$) & $0.0402$ &$0.4437$& $1.2797$ & $2.1952$ & $3.1451$ & $4.1073$ & $5.0763$ & $6.0094$& $6.9258$ &$7.8056$ \\
  \cline{2-12}
       & Im($\om$) & $0.2311$ & $-0.0581$ & $-0.5066$ & $-0.9778$ & $-1.5873$ & $-2.3222$ & $-3.1731$ & $-4.1324$ & $-5.1931$ &$-6.3485$ \\
  \hline
  $L=1.96$ & Re($\om$) & $0.0407$ & $0.6464$ & $1.7115$ & $2.8367$ & $3.9965$ & $5.1664$ & $6.3280$ & $7.4656$ & $8.5658$ &\\
  \cline{2-11}
       & Im($\om$) & $0.1136$ & $-0.2701$ & $-0.7109$ & $-1.3015$ & $-2.0632$ & $-2.9793$ & $-4.0383$ & $-5.2309$ & $-6.5482$ &\\
  \hline
   $L=1.94$ & Re($\om$) & $0.0665$ & $1.5676$ & $2.8389$ & $4.1304$ & $5.4067$ & $6.6680$ & $7.9068$ &  & & \\
  \cline{2-9}
       & Im($\om$) & $-0.1366$ & $-0.6918$ & $-1.0962$ & $-1.8305$ & $-2.7393$ & $-3.7939$ & $-4.9829$ & & & \\
  \hline
\end{tabular}
\end{center}
\end{table}

In Table \ref{tablereim} we list the quasinormal modes shown in
Fig.2. One can see that the imaginary part of the first node becomes
positive when $L=1.96$ and $L=2$, while it is negative if $L=1.94$.
This manifests  that the perturbation becomes unstable when $L$
changes from $1.94$ to $L=1.96$. The critical value is $L=1.951$,
where $\om=0$.

From the left panel of Fig.~\ref{qL} we can  see that when $q=0$,
there are no critical values for $L$ which makes $\om=0$. This
implies that for the neutral fermion perturbation, the black hole is
always stable and the perturbation always decays exponentially.
This is indeed observed in the quasinormal modes calculation as
shown in Table \ref{tablereimwt}, where we see that the imaginary
part of the quasinormal modes of the fermion perturbation is always
negative in the case of $q=0$.

\begin{figure}[htb]
 \includegraphics[scale=0.74]{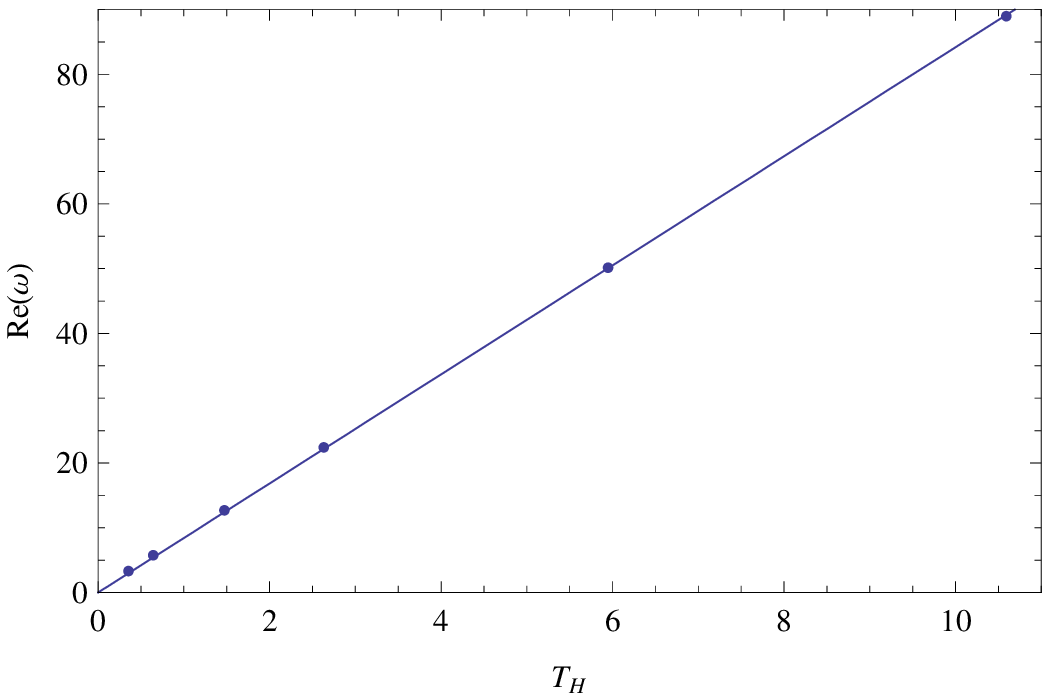}
\includegraphics[scale=0.76]{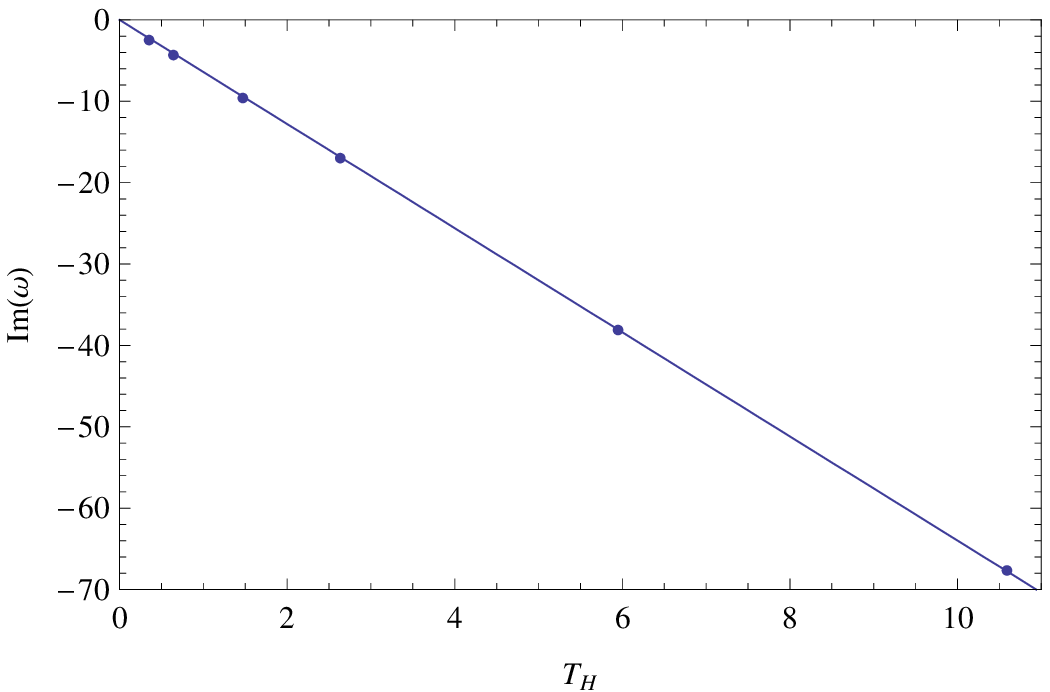}\\
 \includegraphics[scale=0.74]{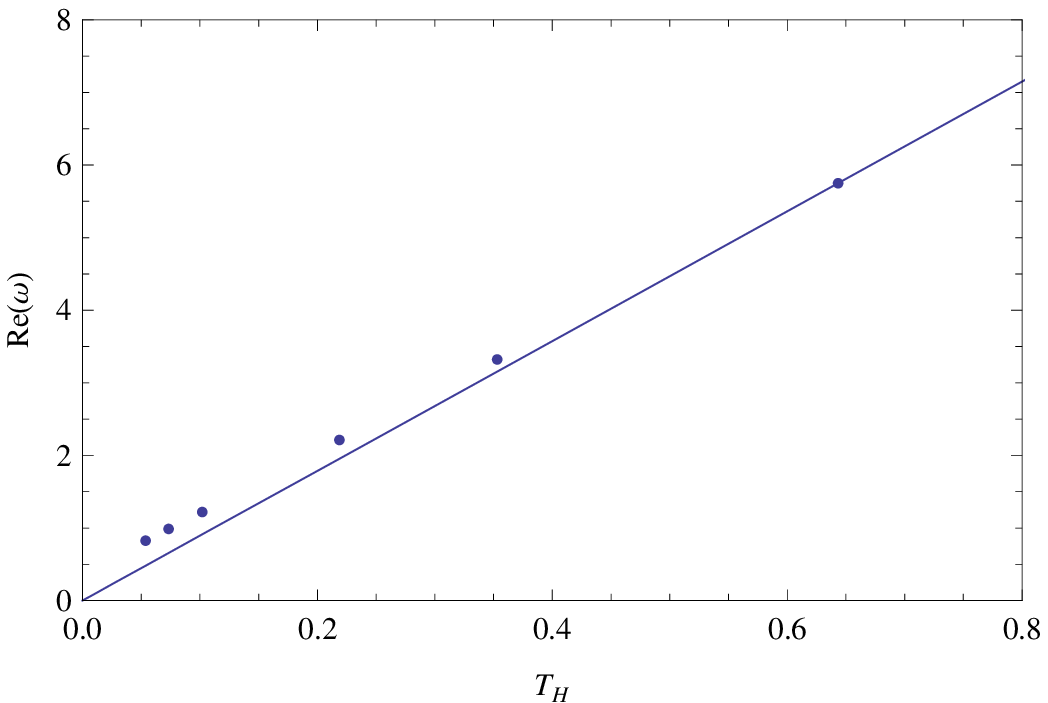}
  \includegraphics[scale=0.74]{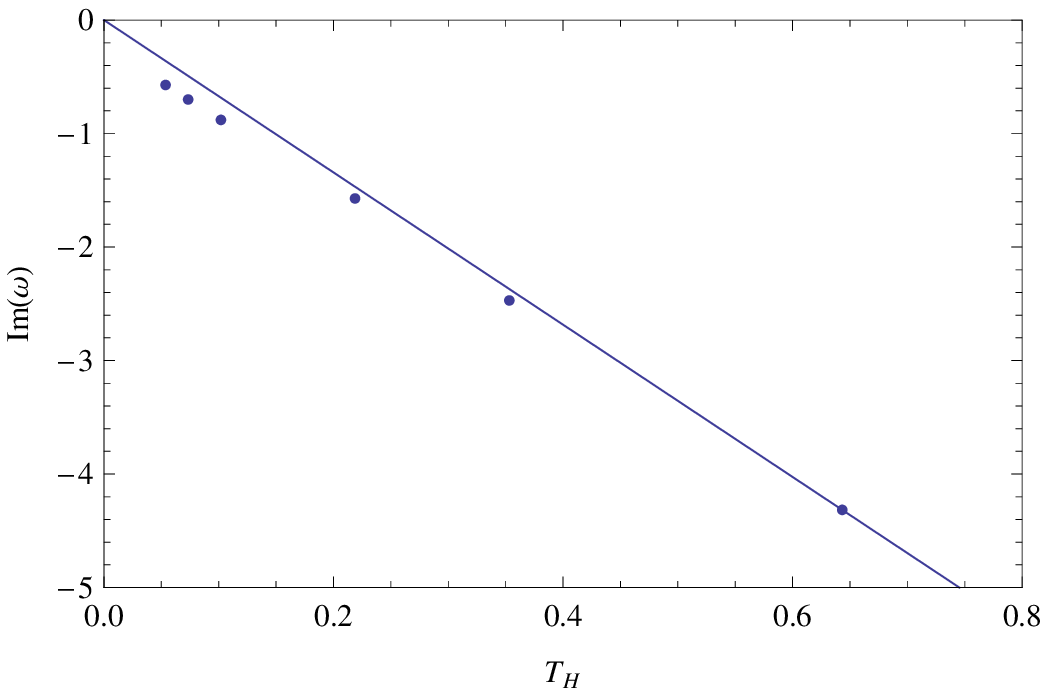}
\caption{\label{reimwt} Quasinormal modes of the first node in the
case of  $q=0$.  The real part versus high temperature (Up-Left);
the imaginary part versus high temperature(Up-Right); the real part
 versus low temperature (Down-Left); the imaginary part versus low
 temperature (Down-Right).}
\end{figure}

\begin{table}
\caption{\label{tablereimwt} Real and imaginary parts of the
quasinormal modes in the case of $q=0$ with different temperature
shown in Fig.~\ref{reimwt}.}
\begin{center}
\begin{tabular}{|c|c|c|c|c|c|c|c|c|c|c|}
  \hline
  $T_H$     & $0.0537$ & $0.0733$ & $0.1019$ & $0.2188$ & $0.3531$ & $0.6432$ & $1.4721$ & $2.6326$ & $5.9484$ &
  $10.5904$
  \\\hline
   Re($\om$) & $0.8253$ & $0.9856$ & $1.2209$ & $2.2134$ & $3.3206$ & $5.7491$ & $12.6862$ & $22.3972$ & $50.1418$ &
   $88.9839$
   \\\hline
  Im($\om$) & $-0.5725$ & $-0.6992$ & $-0.8804$ & $-1.5724$ & $-2.4708$ &$-4.3155$ & $-9.5919$ & $-16.9810$ & $-38.0943$
  & $-67.6533$ \\
  \hline
\end{tabular}
\end{center}
\end{table}
Fig.~\ref{reimwt} shows the behavior of the quasinormal modes with respect to the black hole temperature. We can see from Fig.~\ref{reimwt} that
in the high temperature regime, both the real part and imaginary part of the quasinormal modes have a linear relation to the temperature of the
black hole, which is similar to the result given in~\cite{jing}, where the quasinormal modes of a neutral massless fermion are numerically
calculated in a RN-AdS black hole with a spherical horizon (namely the case with $k=1$ in (\ref{metric})). The linear behavior observed for the
fermonic perturbation agrees to the finding for scalar perturbation \cite{horo, wang}. By numerical fitting, we find that the quasinormal
frequencies behave like \be\label{linear} Re(\om)\approx8.4193T_H,\quad Im(\om)\approx-6.3981 T_H, \ee
 where $T_H$ is the black hole temperature given in (\ref{temp}).
 In the high temperature regime, Ref.~\cite{jing} gives a relation
 (see eq.~(8) in \cite{jing})
 for the case of Schwarzschild-AdS black hole with a spherical
 horizon
 \be
 \label{sch}
 Re(\om)=8.367T,\quad Im(\om)=-6.371T.\ee
 We can see from (\ref{linear}) and (\ref{sch}) that their behaviors
 are quite similar to each other, except the differences in the numerical factors caused by the value of the nonzero
 black hole charge $Q$ we considered here.

In the low temperature regime, on the other hand, one can see that
both the real and imaginary part obviously deviate from the linear
behavior. This result is also consistent with the one of scalar
fields in \cite{horo,wang}.

\section{Concluding Remarks}
\label{sect:con}

In this paper we  considered the Einstein-Maxwell-Fermion system
with a negative cosmological constant. Such a system has been
intensively studied recently in the holographic fermion liquid model
\cite{lee,liu,zaa,rey,wen}. We studied the stability of the RN-AdS
black hole under the perturbation of the charged fermion by
numerically calculating the quasinormal modes of the perturbation in
the probe limit. It is found that when the temperature of the black
hole is below a critical value, the imaginary part of the
quasinormal modes will become positive, which means that the
perturbation of the fermion will grow exponentially which makes the
black hole unstable; while when the temperature of the black hole is
above the critical value, the perturbation will decay exponentially
and the black hole is stable. Our result indicates that a phase
transition occurs in the system from the high temperature phase,
which is described by the RN-AdS black hole, to a low temperature
phase, which should be described by some stable new black hole
solution. Like the case with charged scalar field, we expect that
the stable new black hole should be a one with nontrivial fermion
condensation.

When the coupling constant $q$ increases, the critical temperature of the black hole becomes higher.  This means that the stronger the coupling
is, the much easier for the phase transition to occur.  In the case of $q=0$, namely for a neutral fermion, we found that the imaginary part of
the quasinormal modes is always negative. This implies that the black hole is  stable  under the perturbation of the uncharged fermion and it
will not cause any phase transition of the black hole. The behavior of the quasinormal modes agrees with that for the scalar perturbation
\cite{horo, wang} and in consistent with fermonic perturbation ~\cite{jing} for spherical background. In high temperature regime, both the real
and imaginary parts of the quasinormal modes are linearly proportional to the temperature of the black hole. In the AdS/CFT correspondence,
$1/$Im$(\om)$ is the time scale for the system approaching to thermal equilibrium of the boundary field theory. This means if we perturb the
thermal field on the boundary, the time for it to approach a thermal equilibrium is proportional to the inverse of the temperature~\cite{horo}.
However the linearity breaks in the low temperature regime.

In the probe limit, we found the marginal stable mode exists with
$\om=0$ of the charged fermion. This strongly indicates there should
exist a stable charged black hole with fermion hair in AdS space. It
would be of great interest to consider the backreaction of the
fermion and to find such a black hole solution in our system, which
is currently under investigation.

\begin{acknowledgments}\vskip -4mm

\end{acknowledgments}
H.Q.Z. would like to thank X. He for sharing her program codes and
helpful discussions. Besides, the gratitude will also go to B. Hu
for lending his computer and useful comments. This work was
supported partially by grants from NSFC, China (No. 10821504 and No.
10975168) and a grant from the Ministry of Science and Technology of
China national basic research Program (973 Program) (No.
2010CB833000).

\end{document}